\documentclass[final,5p,times,sort&compress]{elsarticle}

\usepackage{graphicx}
\usepackage{amssymb}

\usepackage{color}
\usepackage{url}

\journal{Forma}

\begin{document}

\begin{frontmatter}

\title{Survival time of Princess Kaguya in an air-tight bamboo chamber}

\author[AI_add]{Akio Inoue}
\author[HS_add]{\corref{cor1}Hiroyuki Shima}
\address[AI_add]{Faculty of Environmental and Symbiotic Sciences, Prefectural University of Kumamoto, 
3-1-100, Tsukide, Higashi-ku, Kumamoto 862-8502, Japan}
\address[HS_add]{Department of Environmental Sciences, University of Yamanashi, 4-4-37, Takeda, Kofu, Yamanashi 400-8510, Japan}

\cortext[cor1]{To whom correspondence should be addressed:\\ 
Hiroyuki Shima (Email: hshima@yamanashi.ac.jp)}

\begin{abstract}
Princess Kaguya is a heroine of a famous folk tale, as every Japanese knows. She was assumed to be confined in a bamboo cavity with cylindrical shape, and then fortuitously discovered by an elderly man in the forest. Here, we pose a question as to how long she could have survived in an enclosed space such as the bamboo chamber, which had no external oxygen supply at all. We demonstrate that the survival time should be determined by three geometric quantities: the inner volume of the bamboo chamber, the volumetric size of her body, and her body's total surface area that governs the rate of oxygen consumption in the body. We also emphasize that this geometric problem shed light on an interesting scaling relation between biological quantities for living organisms.
\end{abstract}

\end{frontmatter}

\section{Introduction}

Imagine being trapped in a closed room with no window, chimney, or hole in the wall to allow the flow of air. Would you survive for long or remain safe in that situation?

The most serious and imminent danger associated with entrapment in a sealed room is the inevitable development of oxygen deficiency disease. Oxygen is one of the key elements required for the sustenance of life, as experimentally confirmed in the 1770's by Joseph Priestley \cite{Priestley}. The author demonstrated that mice could survive in closed containers as long as they contained plants that emit oxygen through photosynthesis; otherwise, the mice could not survive without the plants. This may be true for all mammals including humans \cite{BBCnews}. By incorporating oxygen into the body, humans create approximately 30 molecules of the energy source adenosine triphosphate (ATP) from one molecule of glucose \cite{NilssonBook,Essential}. Therefore, once all the oxygen in a sealed room is consumed by an entrapped individual, the body can no longer produce energy without which it would be difficult to survive. 
An interesting mammalian exception is a rat that lives underground called the naked mole rat \cite{StorzScience}. In the anoxic state, this rat generates energy for survival by a mechanism that is different from ordinary oxygen-based energy production, and, consequently, can survive for 18 minutes in an oxygen-deprived state and does not experience considerable cellular damage \footnote{It was reported in Ref.~\cite{ParkScience} that in emergencies, the naked mole rat has been reported to increase the production of the saccharide fructose in its body, which suggests that it uses fructose, instead of glucose, as an energy source for the survival of relevant tissues such as the brain and heart.}. Of course, it would not be feasible to ``recreate" all mammals as naked mole rats to avoid the risk of oxygen deficiency. Therefore, let us consider how long a human being would survive in a sealed room with no external oxygen supply.

\begin{figure}[ttt]
\centering
 \includegraphics[width=7.5cm]{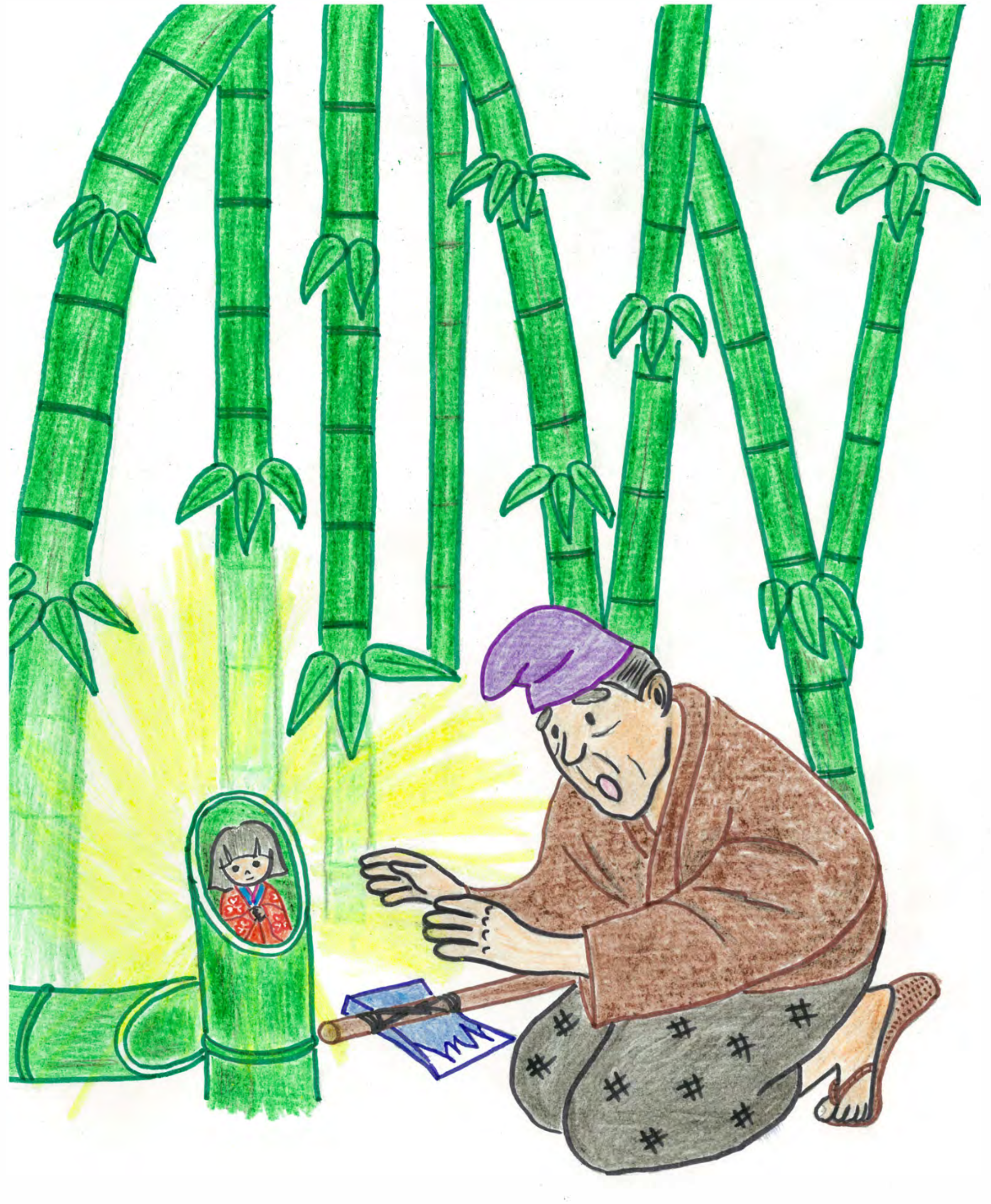}
\caption{Cartoon of the situation where a girl was discovered by an old man. The folk tale states that the old man found a small girl sitting inside a bamboo chamber that shined like gold. He thought the girl must be a god damsel and, so, he then took her home, named her Princess Kaguya, and nurtured her.}
\label{fig_01}
\end{figure}

In the present article, we proposed a Gedanken experiment that provides a strategy for developing a viable solution to the question and problem posed. Our argument is based on a fictional mysterious girl ``Princess Kaguya", who is the heroine of one of the oldest and most famous folk tales in Japan \cite{McCulloughBook}. In the tale, an old man working as bamboo cutter discovered a miniature girl inside a glowing bamboo shoot \cite{quot1}; see Fig.~\ref{fig_01}. Believing her to be a divine presence, he and his wife decided to raise her as their own child. 
When the girl came of age \cite{quot2}, she is conferred with the formal name ``Princess Kaguya", which incorporates the Japanese term Kaguya derived from the light and life she radiated. The legendary story of Princess Kaguya, which has been handed down through generations \cite{LucasBook}  was animated in 2013, released worldwide, and was finally nominated for a 2015 Academy Award for a feature length animation film \cite{MoMA}.

\begin{figure}[ttt]
\centering
\includegraphics[width=7.0cm]{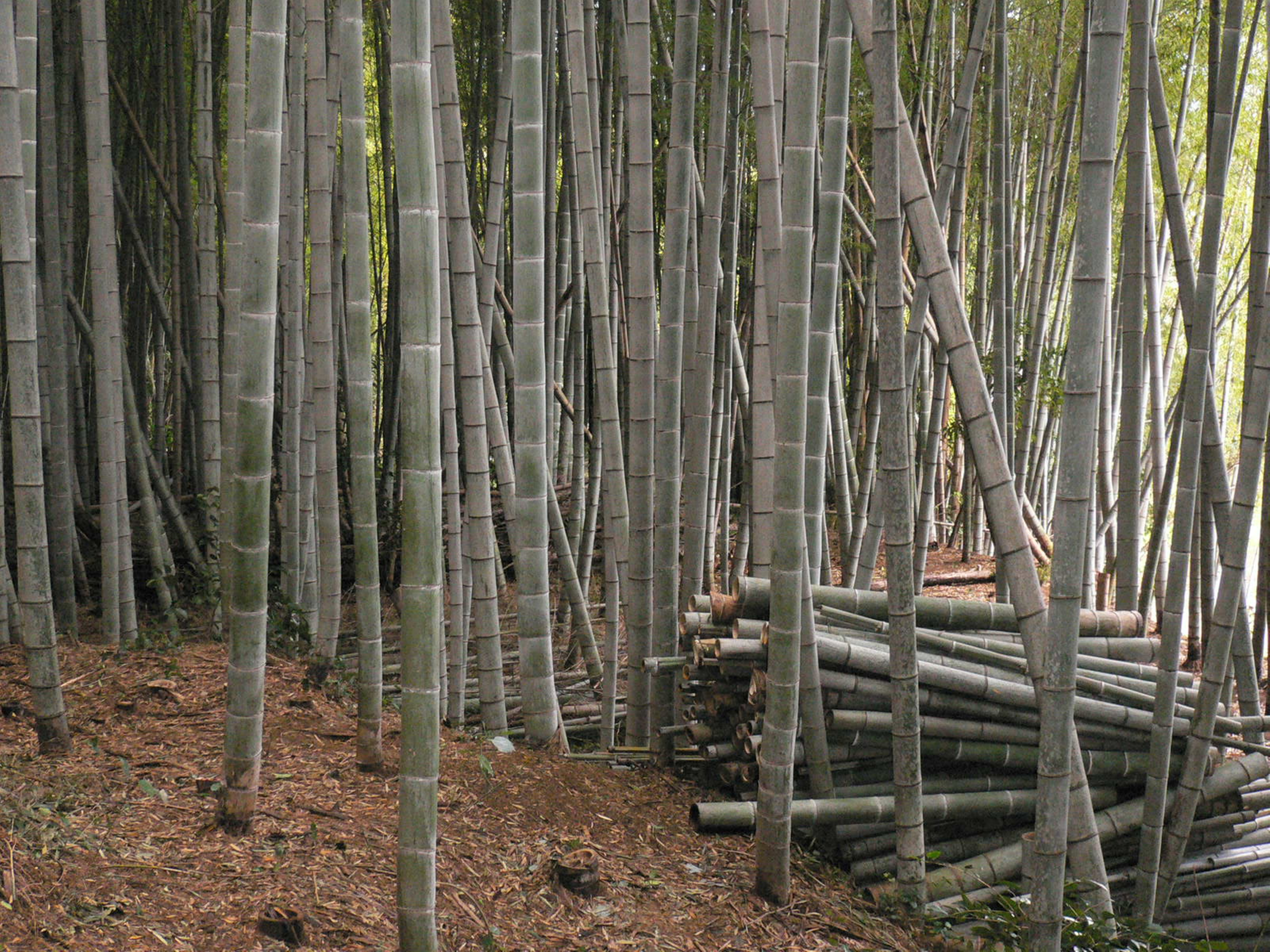}

\vspace*{9pt}
\includegraphics[width=7.0cm]{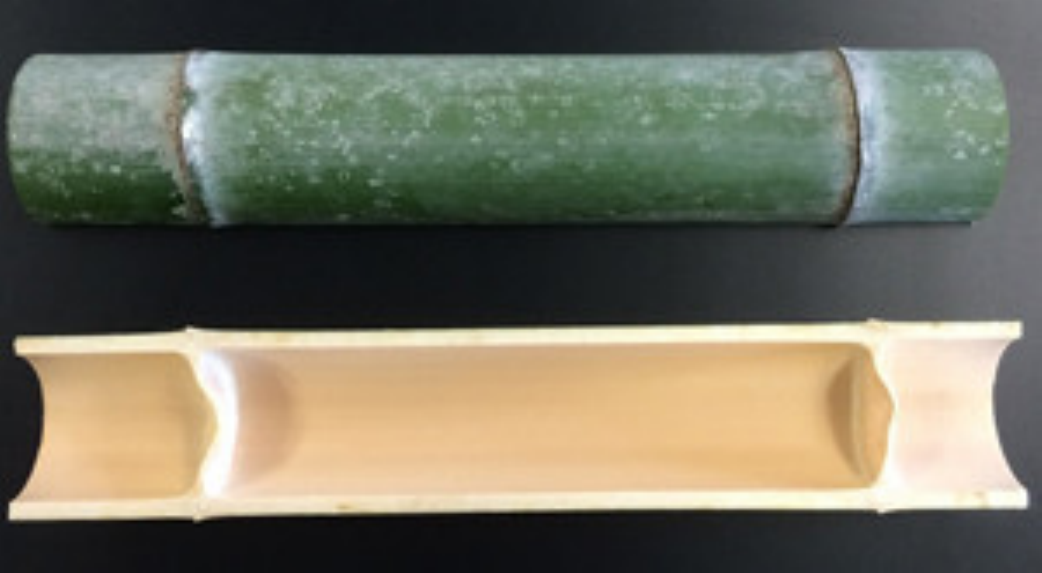}
\caption{(Left) Photo of a bamboo forest. (Right) Cross section of a bamboo culm segment. In the culm, a long cavity inside the cylindrical woody section is divided into numerous chambers by stiff diaphragms.
Reprinted from Ref.~\cite{ShimaPRE2016} with permission.}
\label{fig_02}
\end{figure}

However, we were not concerned with the entire legendary story, but we focused on her entrapment in the sealed space. As mentioned above, Princess Kaguya was assumed to have been discovered inside the stalk of a glowing bamboo plant (see Fig.~\ref{fig_02}), which was completely air-tight. This poses the question of how long Princess Kaguya had survived in the bamboo chamber. We will see below that solving this problem may facilitate our understanding of the interplay between metabolism and respiration \cite{DavidovitsBook,CampbellBook}. In addition, it will be revealed that this problem can be easily extended to situations on different scales, and is not limited to a ``miniature girl situation";  it thus provides an excellent intellectual exercise for considering the allometric relationship of living matter with respect to oxygen.

\section{Assumed condition}

Our discussion will be based on the hypotheses enumerated below.

\begin{itemize}
\item
The space where Princess Kaguya was trapped had cylindrical shape and was a completely sealed room with no inflow and outflow of air. 

\item
The height of the Princess is assumed to be 9.10 cm \cite{McCulloughBook} and not to change while she was trapped.

\item
The Princess remained sitting quietly in the space until her destined discovery by the old bamboo cutter.

\item
Her body produced energy only by breathing the oxygen inside the enclosed space.
\end{itemize}

\section{Way of thinking}

\subsection{Viable time of Princess Kaguya}

\begin{figure*}[ttt]
\centering
  \includegraphics[width=15.5cm]{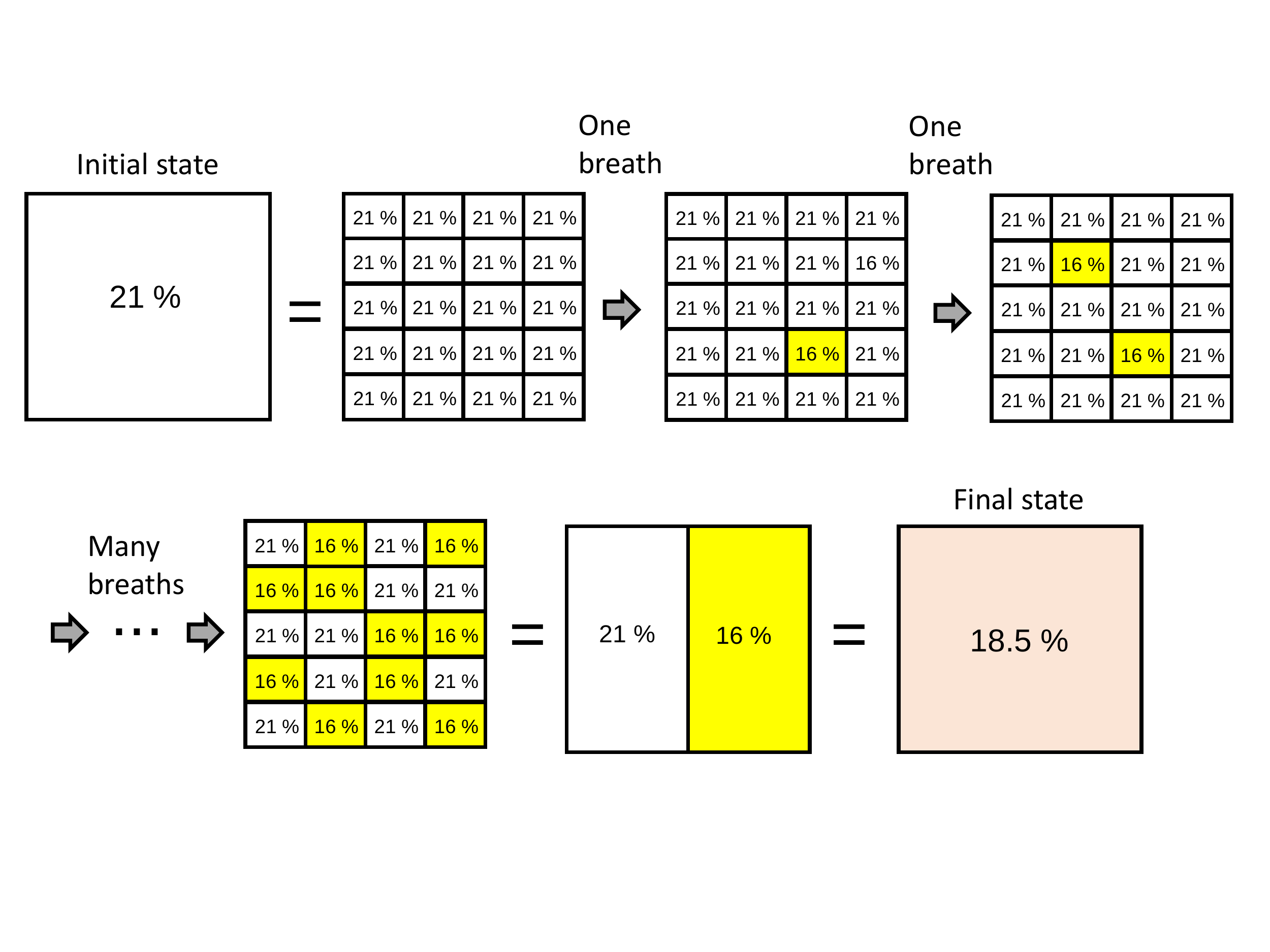}
\caption{Schematic diagram of the time variation in the oxygen concentration inside a bamboo chamber.}
\label{fig_03}
\end{figure*}

Suppose that the inner volume of the bamboo chamber in which the Princess was trapped is $V_{\rm inn}$ [L]. The chamber was filled with air and initially, the oxygen concentration could be assumed to be 21 \%, similar to that of ambient air. For humans in general (including Princess Kaguya), the oxygen concentration of exhaled breath is 16 \%. 
Figure \ref{fig_03} provides a visual illustration;
the outer large square boundary represents the inner chamber that encloses the flesh air with 21 \% in oxygen concentration 
at the initial state.
The total air inside the boundary is divided into those confined in many small square domains,
each of which represents the amount of air that Princess inhales and exhales by one breath.
For simplicity, every exhaled breath does not diffuse but remain in the same position.
\footnote{If the diffusion of exhaled breath is taken into account,
the net oxygen concentration in the inner chamber will decrease with time
and Princess will breathe fast by actively moving diaphragm and muscles needed breathing.
This may cause the increase in the consumed energy per hour and thus
result in a decline of Princess' survival time compared with that obtained under the diffusion-free condition.}
As a result, the oxygen concentration in a small domain 
(highlighted by gray) is displaced by 16 \% following each breath.
Every single breath, Princess inhales flesh air in different small domains in order to intake
a constant amount of oxygen.
Eventually when she breaths half the volume of the total air in the chamber,
a net oxygen concentration wil be 18.5 \% at the final state (see the bottom of Fig.3). 
This value is nearly the same as the lower limit under which oxygen deficiency occurs. 
Taking into account the volume $V_{\rm Pr}$ 
that Princess' body occupies in the chamber, therefore, $(V_{\rm inn}-V_{\rm Pr})/2$ is the total volume of air that the Princess could breathe for survival.

Let $Q_{\rm air}$ be the amount of air 
that Princess intakes per unit time. 
From the above argument,
the viable time $T$ for Princess reads
\begin{equation}
T = \frac{ (V_{\rm inn}-V_{\rm Pr}) }{2 Q_{\rm air}}.
\label{eq_x001}
\end{equation}
In the following discussion, 
we first evaluate 
the Princess-related parameters:
$Q_{\rm air}$ and $V_{\rm Pr}$
in Sections \ref{sec_32} and \ref{sec_33},
and then address the estimation of the bamboo-related parameter $V_{\rm inn}$
in Section \ref{sec_34}.

\subsection{Air consumption rate}
\label{sec_32}

Based on the previous determination, one breath by the Princess constitutes an oxygen intake corresponding to only 5 \% of the total volume of air she inhaled, which could be expressed as follows:
\begin{equation}
Q_{\rm oxy}\; \mbox{[L/h]} = Q_{\rm air}\; \mbox{[L/h]} \times 0.05.
\label{eq_a015}
\end{equation}
Here $Q_{\rm oxy}$ is the amount of oxygen that the Princess intakes per hour, which she uses for energy production.
Hereafter, L/h is the unit of $Q_{\rm oxy}$ and $Q_{\rm air}$.
Human cells use oxygen to create the energy source ({\it i.e.,} ATP) through an oxygen-based chemical reaction to fuel cell functions. 
The amount of energy produced by the intake of 1 L oxygen is 
estimated to be 4.83 kcal/L \cite{Lusk1924}, under the assumption that the respiratory quotient of the
human body\footnote{The respiratory quotient (RQ) is a dimensionless number used in calculating the basal metabolic rate estimated from carbon dioxide production. It is defined as the volumetric ratio of the carbon dioxide removed from the body to the oxygen consumed by the body. The RQ value indicates the kinds of nutrients that are metabolized: values of 0.7, 0.8, and 1.0 indicate that lipids, proteins, and carbohydrates, respectively, are being metabolized. The approximate RQ of a mixed diet is 0.82, which implies that the energy produced following the uptake of 1 L oxygen is 4.83 kcal/L as derived from Ref.~\cite{Lusk1924}.}
is 0.82.
As a result, when the oxygen uptake rate is $Q_{\rm oxy}\; \mbox{[L/h]}$, the following applies:
\begin{equation}
E_{\rm p}\; \mbox{[kcal/h]}
= 4.83\; \mbox{[kcal/L]} \times Q_{\rm oxy}\; \mbox{[L/h]},
\label{eq_a010}
\end{equation}
where $E_{\rm p}$ is the rate of energy production following the intake of oxygen 
at the rate of $Q_{\rm oxy}$. This is the energy consumed by the body for the maintenance of physiological life activities.

\begin{table}[bbb]
  \begin{center}
    \caption{Metabolic rates for selected activities \cite{DavidovitsBook}.}
    \begin{tabular}{|l|c|} \hline
      Activity & Metabolic rate [kcal/(m$^2$ $\cdot$ h)] \\ \hline
      Sleeping & 35 \\
      Sitting upright & 50 \\
      Standing & 60 \\
      Walking & 140 \\
      Running & 600 \\ \hline
    \end{tabular}
  \end{center}
  \label{table_01}
\end{table}

It is interesting to note that the consumption energy by human and many kinds of mammals is proportional to the total body surface area \cite{Robertson1952}. This is primarily because many mammals regulate their body temperature by heat exchange through the skin surface and the respiratory tract to maintain a constant body temperature at an extremely narrow range ({\it e.g.}, 36-37 $^\circ$C for humans). Especially at rest, most of this heat exchange occurs through the skin surface. Therefore, it would be easy to determine there is a high degree of correlation between the body surface area and the energy consumption rate of the human body. The value of the proportionality constant is determined by the degree of active motion of the body, as listed in Table 1.

For instance, when a person is sitting still the energy consumption per hour per surface area
is 50 kcal/(m$^2 \cdot$ h). 
Accordingly, the energy $E_{\rm c}$ consumed per hour can be calculated as follows:
\begin{equation}
E_{\rm c}\; \mbox{[kcal/h]} 
= 50\; \mbox{[kcal/(m$^2 \cdot$ h)]} \times A \; \mbox{[m$^2$]},
\label{eq_a005}
\end{equation}
where $A$ [m$^2$] is the total surface area of the body. In the present study, we assumed that the energy was produced only by breathing and, consequently, $E_{\rm c} = E_{\rm p}$ .

From Eqs.~(\ref{eq_a015})-(\ref{eq_a005}), we obtain
\begin{equation}
Q_{\rm air}\; \mbox{[L/h]} 
= \frac{ 50\; \mbox{[kcal/(m$^2 \cdot$ h)]} \times A\; \mbox{[m$^2$]} }{ 4.83\; \mbox{[kcal/L]} \times 0.05}.
\label{eq_x002}
\end{equation}
Therefore, the remaining task is the determination of the body surface area ($A$) of the Princess, which will be solved in the next subsection.

\subsection{Total surface area of the human body}
\label{sec_33}

Developing methods for estimating the surface area of the human body has long been an important subject in the field of biology, physiology, medicine, and engineering \cite{Verbraecken,Piccirilli}. It is currently accepted that 
the total surface area $A$ [m$^2$] of the human body generally obeys a power law with respect to the height $H$ [m] and weight $M$ [kg] of the body as provided by \cite{DuBois1915}
\begin{equation}
A = 0.202 \times M^{0.425} \times H^{0.725}.
\end{equation}
Let us use this power law to evaluate $A$ of Princess.
First, we remind that the body height of Princess Kaguya was assumed to be
$H= 9.10$ cm \cite{McCulloughBook}. 
Next, we considered the weight of the Princess, which was finally determined by assuming a similar relationship between  $H$ and $M$ based on different scales as explained below. In Japan, the average height ($H_0$) and weight ($M_0$) of 5-years-old girls are
$H_0=107.87$ cm and $M_0=17.6$ kg, respectively \cite{MHLW}.
Assuming the similarity relationship, therefore,
the body weight of the Princess, $M$, is
\begin{eqnarray}
M &=&
M_0 \times \left( \frac{H}{H_0} \right)^3 
= 17.6\; \mbox{[kg]} 
\times
\left( \frac{9.10 \; \mbox{[cm]}}{107.87\; \mbox{[cm]}} \right)^3 
\nonumber \\
&=& 10.6 \times 10^{-3}\; \mbox{[kg]}.
\label{eq_045m}
\end{eqnarray}
In summary, the final calculation for Princess Kaguya is as follows:
\begin{eqnarray}
A &=& 0.202 
\left( 10.6 \times 10^{-3} \right)^{0.425}
\times \left( 9.10  \times 10^{-2} \right)^{0.725} \nonumber \\
&=& 
5.14 \times 10^{-3}\; \mbox{[m$^3$]}. 
\label{eq_x003}
\end{eqnarray}
In addition, the volume of Princess' body $V_{\rm Pr}$ can be deduced
from Eq.~(\ref{eq_045m}) as
\begin{equation}
V_{\rm Pr} = 
10.6\; \mbox{[cm$^3$]},
\end{equation}
where the specific gravity of the Princess's body was assumed to be equal to that of water: 1 g$/$cm$^3$.

\subsection{Inner volume of a bamboo chamber}
\label{sec_34}

\begin{figure}[ttt]
\begin{center}
  \includegraphics[width=7.5cm]{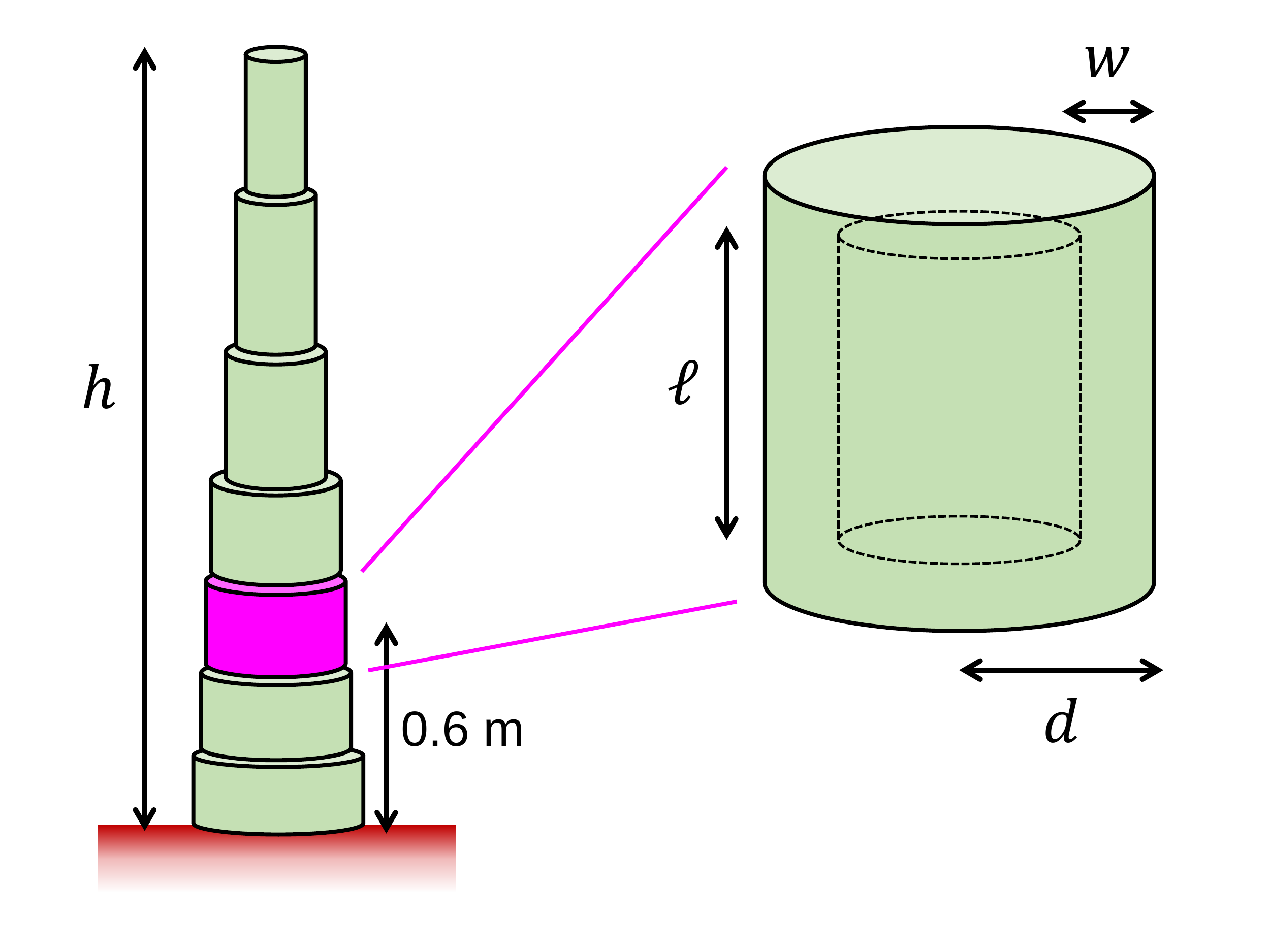}
\end{center}
\caption{A bamboo culm with the total height of $h$
and the internode located 0.6 m above the ground.
The parameters $\ell$, $d$, $w$
used in the text are also defined.}
\label{fig_04}
\end{figure}

If the bamboo chamber in which the Princess was trapped was assumed to be cylindrical with no tapering, the inner volume of the chamber $V_{\rm inn}$ could be calculated as follows:
\begin{equation}
V_{\rm inn} = \frac{\pi}{4} \ell (d-2w)^2,
\label{eq_056v}
\end{equation}
where $\ell$ is the internode length, $d$ is the outer diameter of the cross section, and $w$ is the wall thickness of the bamboo culm. Figure \ref{fig_04} shows the schematics of the three geometric parameters and that of the total height $(h)$ of a bamboo culm. We hypothesize that the Princess was in a chamber with a height of 0.6 m
from the ground. This would be a reasonable height for an elderly man to cut with an axe or sickle while stooping. Furthermore, since the internode length decreases as it approaches the ground from above \cite{Inoue2017}, those located lower than 0.6 m in altitude would tend to have had insufficient volume to enclose the Princess with a height of 9.10 cm.

\begin{table}[bbb]
  \begin{center}
    \caption{Measurement data of the bamboo culm sizes.}

  \begin{tabular}{|c|c|c|} \hline
  Species     & {\it Phyllostachys}  & {\it Phyllostachys} \\ [-4pt]
              & {\it bambusoides}    & {\it nigra} \\ [-4pt]
              & Sieb. et Zucc.       & var. {\it henonis} \\  [-4pt]
              & (Madake)             & (Hachiku) \\ \hline
  $h$ [m]     & 13.9 & 8.3 \\
  $\ell$ [cm] & 20   & 20  \\
  $d$ [cm]    & 6.2  & 3.8 \\
  $w$ [cm]    & 0.74 & 0.65 \\ \hline
  \end{tabular}

    \label{table_02}
  \end{center}
\end{table}

To quantify $V_{\rm inn}$ of the internode chamber 
depicted in Fig.~\ref{fig_04}, 
we had to specify the species of bamboo in which the Princess was trapped. There were two possible candidates of bamboo species, which are listed in Table 2 and commonly called ``Madake" and ``Hachiku" in Japan \cite{BessBook}.

These two species have existed in Japan for at least 800 years since they were domesticated in the late 9th century to the early 10th century when the tale of Princess Kaguya is thought to have been written.
Most of the other bamboo species in Japan can be excluded as potential candidates because they were introduced in Japan only a few centuries ago \cite{Suzuki} or they would not have had a sufficient volume to confine the Princess with a 9.10 cm height. Based on this background, we focused on the two bamboo species that have existed from ancient times (see Table \ref{table_02}), and actually personally collected several samples from the forest (similar to what the bamboo cutter did) to measure the $\ell$, $d$, and $w$ values using Eq.~(\ref{eq_056v}). The internode was measured at a height approximately 0.6 m from the ground.

Using the measurement values listed in Table \ref{table_02} we used the following equation for the subsequent calculation:
\begin{equation}
V_{\rm inn} = 3.5 \times 10^2\; \mbox{[cm$^3$]} \;\; \mbox{for Madake},
\label{eq_x004}
\end{equation}
and
\begin{equation}
V_{\rm inn} = 98\; \mbox{[cm$^3$]} \;\; \mbox{for Hachiku}.
\label{eq_x004}
\end{equation}

\section{Results}

Substituting the results obtained thus far into Eq.~(\ref{eq_x001}), 
we arrived at the conclusion that the viable time 
for Princess Kaguya could be calculated as follows:
\begin{equation}
T = 0.16\; \mbox{[h]} = 9.6\; \mbox{[min]} \;\; \mbox{for Madake},
\label{eq_x005}
\end{equation}
and
\begin{equation}
T = 0.041\; \mbox{[h]} =2.5\; \mbox{[min]} \;\; \mbox{for Hachiku}.
\label{eq_x006}
\end{equation}
These are the solutions to our questions on the viable time for human survival in the bamboo chamber.

It should be noted that in both cases above, the viable time was found to be less than 10 min, which appears to be surprisingly short considering that the Princess was allegedly rescued by the old man 
who was passed by the bamboo she was trapped in by chance. In fact, the situation allowed of no delay; the life of the Princess would have been in danger if the old man had not rescued her promptly. This observation would create the impression that she was an extremely fortunate or blessed girl. Alternatively, it could be proposed that she was an ``ancestor" of the naked mole rat and was endowed with extreme tolerance against oxygen deprivation; the acceptable notion would ultimately be based on the readers' discretion or decision.

\section{Biophysical scaling relation}

From the viewpoint of biophysics,
the shortness in $T$ given by Eqs.~(\ref{eq_x005}) and (\ref{eq_x006})
is regarded as an important consequence
of an allometric relation deduced from Eq.~(\ref{eq_x001}).
Note that in the right-hand side of Eq.~(\ref{eq_x001}),
the numerator ({\it i.e.}, $V_{\rm inn}-V_{\rm Pr}$) scales in cubic manner 
to the length dimension,
and the dominator ({\it i.e.,} $Q_{\rm air}$) scales in square manne,, respectively. Therefore, the viable time $T$ was proportional to the characteristic length of the system. Specifically, the smaller the body size and trapping space, the shorter the survival time of the live organism therein, although the ratio of the both sizes remains unchanged
({\it i.e.} even when the geometric similarity holds). In this context, we considered the viable time problem to be an allometric problem, which would be instructive in acquiring a better understanding of the scaling relation between biological quantities.

\begin{figure}[ttt]
\begin{center}
  \includegraphics[width=4.5cm]{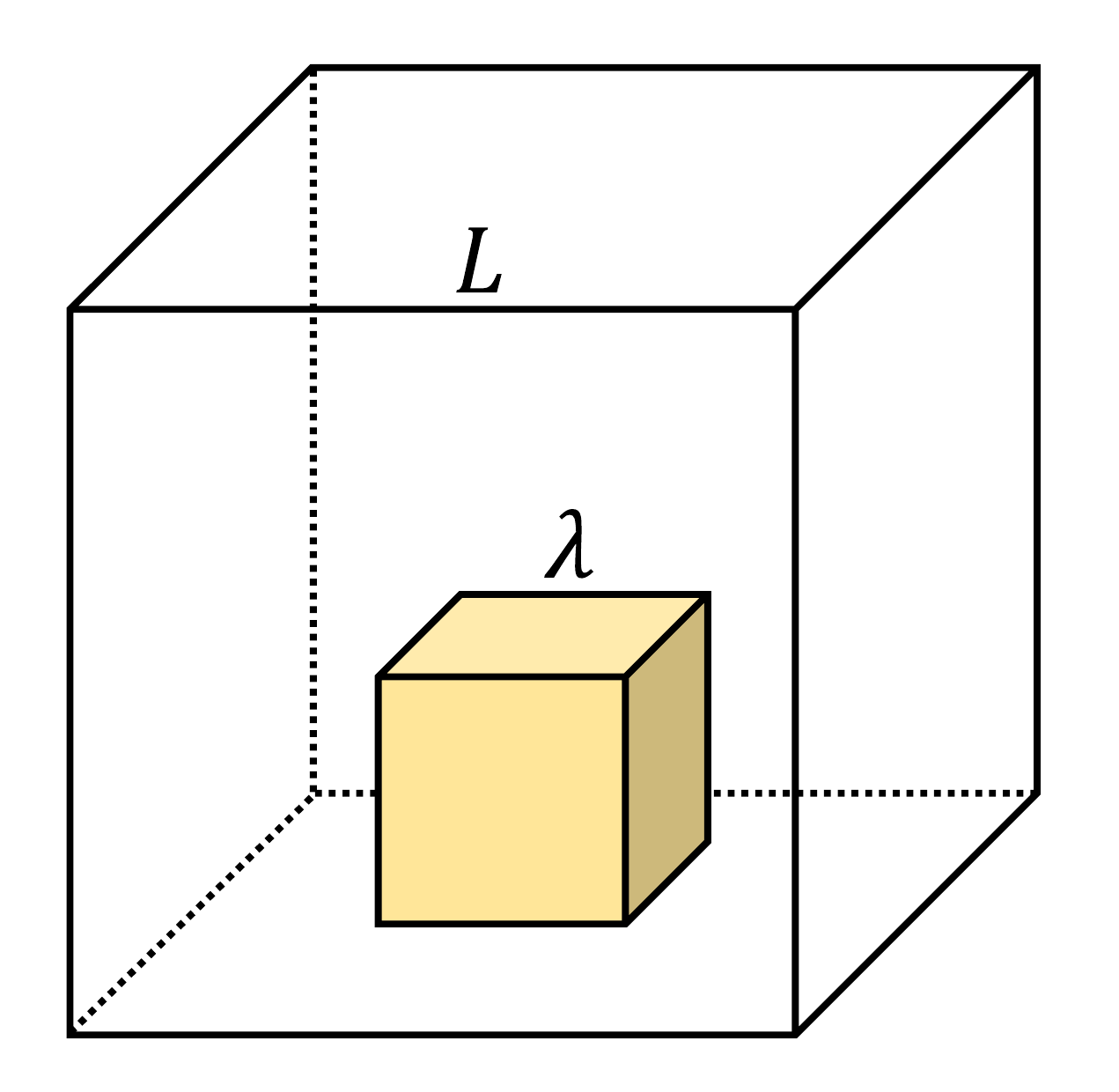}
\end{center}
\caption{Cubic-shaped closed space with a side length of $L$ that contains a cubic-shaped living organism with $\lambda$ as the linear dimension.}
\label{fig_05}
\end{figure}

The above discussion can be clarified by the information illustrated in Fig.~\ref{fig_05}, which presents a simplified situation in which a cube with length of side $\lambda$ is confined in a cubic cavity with length of side $L$. Using the inner cube to simulate a living organism, we could then obtain the equation

\begin{equation}
V_{\rm inn} = L^3, \quad V_{\rm Pr}=\lambda^3, \quad Q_{\rm air}\propto \lambda^2,
\end{equation}
which implied that
\begin{equation}
T\propto \frac{L^3 - \lambda^3}{\lambda^2}.
\end{equation}
If $L/\lambda$ is fixed as a constant $c$, the following equation is obtained:
\begin{equation}
T\propto \frac{\lambda^3 (c^3-1)}{\lambda^2} \propto \lambda.
\label{eq_111}
\end{equation}
Equation (\ref{eq_111}) reveals that the viable time is proportional to 
the characteristic body length $\lambda$ when the geometric similarity $L/\lambda \equiv c$ is satisfied.

\section{Summary}
In this study, we estimated the time for which Princess Kaguya could have survived in the completely airtight bamboo chamber. In the analysis, we used several basic concepts associated with the metabolic rate and geometric relationship of the human body, which are both frequently encountered
in the ordinary course of biophysics. Based on the assumed sizes of the 
Princess' body and bamboo chamber, we concluded that the duration was less than 10 min, which was shorter than what would be intuitively expected. The short duration is attributable to the allometric scaling, which implies that the viable time duration is linearly scaled to the characteristic length scale of the system.

As a final remark,
we emphasize that the problem addressed is not limited to 
a fictitious ``miniature-girl" situation,
but can be extended easily to
different situation in the body size, the amount of activity,
and the number of persons trapped in a sealed space.
The following questions could be reasonably proposed for further 
investigations,
which will provide good exercises for the study of scaling relation between biological quantities.

\begin{itemize}

\item
Consider a situation where a  normal-sized person is trapped in an airtight safe made of steel. Using a self-determined size for the safe, estimate the time that this person had to breathe.

\item 
Discuss a solution similar to that described above, in which 100 spectators are trapped in an airtight concert hall.

\item
Consider the breathing-time problem for a normal-sized person who moves actively in a sealed room using the numeric data listed in Table 1.

\end{itemize}

\section*{Acknowledgments}
The authors acknowledge 
Prof.~Eiichi Yoshimura for helpful comments
on the estimation of basal metabolic rate, and
Prof.~Motohiro Sato for constant discussions about bamboo science.
This work was supported by JSPS KAKENHI Grant Numbers 
JP 25390147 and 16K14948.


\begin{thebibliography}{99}


\bibitem{Priestley}
Priestley J 1775
An account of further discoveries in air
{\it Phil. Trans.} {\bf 65} 384-394

\bibitem{BBCnews}
BBC News 2011 {\it Scientist starts plant-only oxygen test at Eden Project}
(16 September 2011) http://www.bbc.com/news/uk-england-14944947

\bibitem{NilssonBook}
Nilsson GE,
Chapter 1 in {\it Respiratory Physiology of Vertebrates},
(Cambridge University Press, 2010)

\bibitem{Essential}
Alberts B {\it et al.,} Chapter 14 in {\it Essential Cell Biology}, 3rd ed., (Garland Science, 2009)

\bibitem{StorzScience}
Storz JF and McClelland GB 2017
Rewiring metabolism under oxygen deprivation
{\it Science} {\bf 356} 248-249

\bibitem{ParkScience}
Park TJ {\it et al.} 2017
Fructose-driven glycolysis supports anoxia resistance in the naked mole-rat
{\it Science} {\bf 356} 307-311

\bibitem{McCulloughBook}
McCullough HC,
{\it Classical Japanese Prose: An Anthology},
(Stanford Univ Press, 1991)

\bibitem{quot1}
From Ref. \cite{McCulloughBook}: {\it ``Once upon a time, there was an old bamboo cutter who went into the mountains and fields, cut bamboo, and put the stalks to all kinds of uses. $\ldots$ Now it happened that one stalk of bamboo shone at the base. $\ldots$ When he examined it, he saw a dainty little girl, just three inches tall, sitting inside."}

\bibitem{quot2}
From Ref. \cite{McCulloughBook}: {\it ``He put here in his hand, took her home, and gave her to his wife to rear. She was the cutest in the world. $\ldots$ The child shot up swiftly while they cared for her. By the time she was three months old, she was as big as an adult."}

\bibitem{LucasBook}
Lucas S {\it Bamboo} (Reaktion Books Ltd., 2013)

\bibitem{MoMA}
``The Tale of Princess Kaguya" (Directed by Isao Takahata, 2014), The Museum of Modern Art
\verb|https://www.moma.org/calendar/events/714|

\bibitem{ShimaPRE2016}
Shima H, Sato M, and Inoue A 2016
Self-adaptive formation of uneven node spacings in wild bamboo
{\it Phys.~Rev.~E} {\bf 93} 022406

\bibitem{DavidovitsBook}
Davidovits P, Chapter 11 in 
{\it Physics in Biology and Medicine}, 4th ed.
(Academic Press, 2012)

\bibitem{CampbellBook}
Campbell GS and Norman JM, Chapter 13 in
{\it An Introduction to Environmental Biophysics}, 2nd ed.
(Springer, 1998)


\bibitem{Lusk1924}
Lusk G 1924
Analysis of the oxidation of mixtures of carbohydrate and fat
{\it J. Biol. Chem.} {\bf 59} pp.41-42.


\bibitem{Robertson1952}
Robertson JD and Reid DD 1952
Standards for the basal metabolism of normal people in Britain
{\it Lancet} {\bf 1} 940-943

\bibitem{Verbraecken}
Verbraecken J, de Heyning PV, De Backer W, and Gaal LV 2006
Body surface area in normal-weight, overweight, and obese adults. A comparison study
{\it Metabolism} {\bf 55} 515-524

\bibitem{Piccirilli}
Piccirilli M, Doretto G, and Adjeroh, D 2017
A framework for analyzing the whole body surface area from a single view
{\it PLOS ONE} {\bf 12} e0166749

\bibitem{DuBois1915}
Du Bois D, and Du Bois EF 1915
The measurement of the surface area of man
{\it Arch.~Intern.~Med.} {\bf 15} 868-881


\bibitem{MHLW}
Report on National Health and Nutrition Survey 2014: 
The Ministry of Health, Labor and Welfare.
\verb|http://www.mhlw.go.jp/toukei/youran/indexyk_2_1.html|

\bibitem{Inoue2017}
Inoue A, Tochihara S, Sato M, and Shima H 2017
Universal node distribution in three bamboo species ({\it Phyllostachys} spp.)
{\it Trees} {\bf 31} 1271-1278


\bibitem{BessBook}
Bess NM {\it bamboo in japan} (Kodansha Int'l Ltd., 2001).

\bibitem{Suzuki}
Suzuki S and Nakagoshi N
{\it Sustainable management of Satoyama bamboo landscape in Japan}. 
In: {\it Landscape ecology in Asian cultures}
(eds. Hong SK {\it et al.}) pp.211-220 (Springer, 2011).


\end{thebibliography}
\end{document}